\documentclass[onecollarge,natbib]{svjour2}
\bibpunct{[}{]}{,}{n}{}{,} 
\smartqed  
\usepackage{graphicx}
\newcommand{\be}{\begin{equation}}
\newcommand{\ee}{\end{equation}}
\newcommand{\ben}{\begin{eqnarray}}
\newcommand{\een}{\end{eqnarray}}

\newcommand{\nn}{\nonumber}
\journalname{Few-Body Systems (APFB2011)}
\begin{document}

\title{\boldmath Scattering of Two Spinless Particles in 3D Formulation
with Coulomb Admixtures}


\author{Fahmi Maulida     \and
        Imam Fachruddin} 

\authorrunning{F. Maulida \and I. Fachruddin } 

\institute{F. Maulida \at
              Departemen Fisika, Universitas Indonesia, Depok 16424, Indonesia\\
              \email{cipangimpenan@gmail.com}
           \and
           I. Fachruddin \at
              Departemen Fisika, Universitas Indonesia, Depok 16424, Indonesia}

\date{Received: date / Accepted: date}

\maketitle

\begin{abstract}
Scattering of two spinless charge particles for simple forces
including coulomb admixtures is calculated without partial wave
decomposition. The coulomb interaction being taken is of the type
of screened coulomb potential. For the forces range are not infinite,
the standard scattering theory is applied. The differential and
total cross section is calculated and coulomb effects are shown.
\keywords{3D Technique \and Coulomb Scattering}
\end{abstract}

\section{Introduction}
\label{intro}
Coulomb scattering and scattering with coulomb admixtures have long been challenging problems.
This is due to the infinite range of the Coulomb interaction. And although Coulomb interactions
in practice are always screened, so that conventional nonrelativistic scattering theory can be
applied\cite{taylor,semon}, the effective interaction range is still large. Using the standard
partial-wave (PW) technique this means one has to sum up very many partial-wave, which can lead
to some difficulties in numerical realization, especially if at the same time one also considers
processes in higher energy region.

As an alternative to the PW technique is the so called three-dimensional (3D) approach. Since
in 3D calculations the basis states are not expanded into partial waves, there is no problem with
partial wave summation, since a priori all partial waves are included. Thus, here we propose to
use a 3D approach to calculate scattering with Coulomb admixtures. As an example in this first step,
we choose a simple system, that is scattering of two bosons. As interaction we take that of
the type of the Malfliet-Tjon\cite{mtj}. We choose to considered a screened Coulomb potential,
with the screening factor being an exponential function. We apply a 3D technique for two boson
scattering formulated in Ref.~\cite{thomas}.

In Section \ref{form} we shortly show the 3D formulation, which is given in details in Ref.~\cite{thomas}.
In Section \ref{redisc} we show some calculations for differential and total cross sections, where we
can see some Coulomb effetcs. We summarize in Section \ref{summary}.

\section{Formulation}
\label{form}
The Lippmann-Schwinger equation for the $T$-matrix elements based on the 3D basis states 
$|\textbf{q}\rangle$ is given as 
\be
T(\textbf{q}',\textbf{q})=V(\textbf{q}',\textbf{q})+\lim_{\epsilon \rightarrow 0}\int{d^3q''
V(\textbf{q}',\textbf{q}'')\frac{1}{\frac{{q}^2}{2m}+i\epsilon-\frac{{q}''^2}{2m}}}T(\textbf{q}'',\textbf{q}) \, , 
\label{eq:T}
\ee
with $\textbf{q}$ being the relative momentum, $m$ the reduced mass of the system, $V(\textbf{q}', \textbf{q})$ 
and $T(\textbf{q}', \textbf{q})$ the potential and the $T$-matrix elements. Since the potential and the $T$-matrix elements 
are scalar functions, i.e. 
\ben
V(\textbf{q}',\textbf{q})&=&V(q',q,\hat{\textbf{q}'}\cdot\hat{\textbf{q}})\\
T(\textbf{q}',\textbf{q})&=&T(q',q,\hat{\textbf{q}'}\cdot\hat{\textbf{q}}) \, , 
\een
Eq.~(\ref{eq:T}) finally becomes 
\be
T(q',q,x')=V(q',q,x')
+\lim_{\epsilon \rightarrow 0}\int_0^\infty{dq''}q''^2\int_{-1}^1{dx''} v(q',q'',x',x'')
\frac{1}{\frac{{q}^2}{2m}+i\epsilon-\frac{q''^2}{m}}T(q'',q,x'')\, , 
\label{eq:1}
\ee
where $x\;'=\hat{\textbf{q}}\;'\cdot\hat{\textbf{q}}$, $x\;''=\hat{\textbf{q}}\;''\cdot\hat{\textbf{q}}$,
and 
\be
v(q',q'',x',x'')=\int{d\varphi''}V(q',q'',\hat{\textbf{q}}'\cdot\hat{\textbf{q}}'')\, . \label{vvphi}
\ee

We choose for the short-range interaction those of the type of the Malfliet-Tjon potential,
\be
V(r)=V_R\frac{exp[-\mu_Rr]}{r}-V_A\frac{exp[-\mu_Ar]}{r}\, , 
\ee
with $V_R$,  $V_A$, $\mu_R$, and $\mu_A$ being some parameters. For the Coulomb interaction 
we take the screened one, with the screening factor being an exponential function,  
\be
V_C(r)=V_C\frac{exp[-\mu_Cr]}{r} \, , 
\ee
where $V_C$ and $\mu_C$ are the Coulomb parameters. The full interaction in momentum representation is, 
therefore,  
\be
V(\textbf{q}',\textbf{q})=\frac{1}{2\pi^2}\left\lbrace\frac{V_R}{(\textbf{q}'-\textbf{q})^2+\mu_R^2}-\frac{V_A}{(\textbf{q}'-\textbf{q})^2+\mu_A^2}\
+\frac{V_C}{(\textbf{q}'-\textbf{q})^2+\mu_C^2}\right\rbrace \, . \label{qqq}
\ee
For the potential given in Eq.~(\ref{qqq}) the integration in Eq.~(\ref{vvphi}) can be evaluated analitically. We obtain
\ben
v(q',q'',x',x'')&=&\frac{1}{\pi}\Biggl[\frac{V_R}{\sqrt{(q'^2+q''^2-2q'q''x'x''+\mu_R^2)^2-4q'^2q''^2(1-x'^2)(1-x''^2)}}\nn\\
&&-\frac{V_A}{\sqrt{(q'^2+q''^2-2q'q''x'x''+\mu_A^2)^2-4q'^2q''^2(1-x'^2)(1-x''^2)}}\nn\\
&&+\frac{V_C}{\sqrt{(q'^2+q''^2-2q'q''x'x''+\mu_C^2)^2-4q'^2q''^2(1-x'^2)(1-x''^2)}}\Biggr] \, . 
\een

\section{Result and discussion}
\label{redisc}
Here we show some calculations for differential and total cross section. As parameters we take $V_R=7.291$, $\mu_R=613.69$ MeV, 
$V_A=-3.1769$, $\mu_A=305.86$ MeV, and $V_C=0.0073$. 

\begin{figure}
\centering
\input{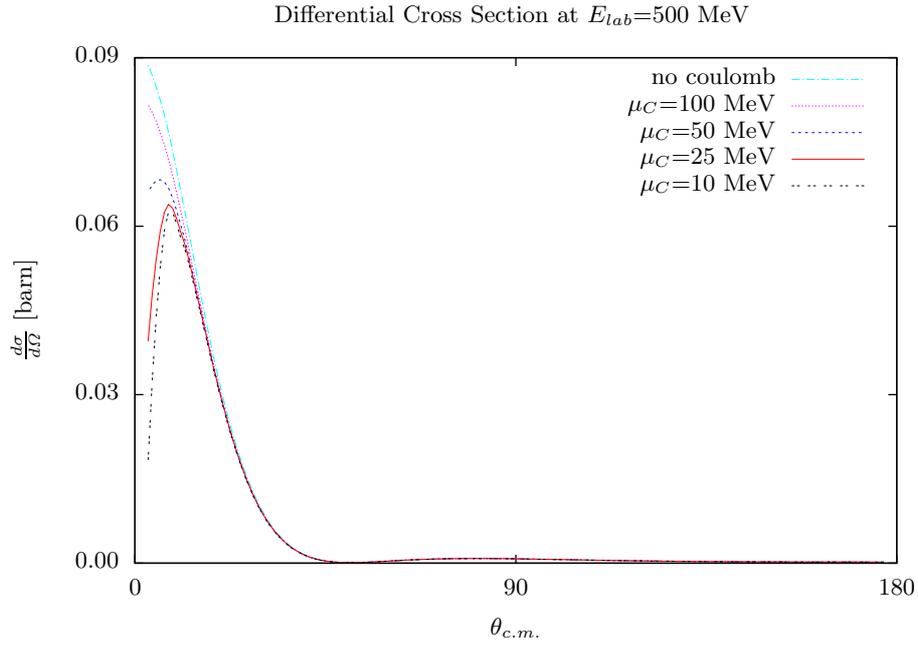}
\caption{(color online) Differential cross section for various range of Coulomb
interaction represented by $\mu_C$.}
\label{fig1}
\end{figure}

First, we observ some Coulomb effects in differential cross section, while varying the range of the Coulomb interaction. 
As large interaction range as well as high scattering energy require more integration points, especially that for the integration 
over the momentum's magnitude, we firstly find the optimal number of integration points for very large Coulomb range 
corresponding to $\mu_C=10$ MeV and energy of 1000 MeV. The obtained number of integration points can then be safely used for 
smaller Coulomb range and lower energies. In Fig.~\ref{fig1} we show differential cross section for $E_{lab}=500$ MeV and various 
Coulomb range represented by the Coulomb parameter $\mu_C$. Coulomb effects are observed in forward direction. Figure \ref{fig1} 
also shows that sufficiently large Coulomb range has to be considered in order to investigate Coulomb effects. For the system being 
considered varying the Coulomb range together with applying the 3D technique leads to no difficulty. We consider that $\mu_C=10$ MeV 
is sufficient to represent some kind of effective Coulomb range. 

\begin{figure}
\centering
\input{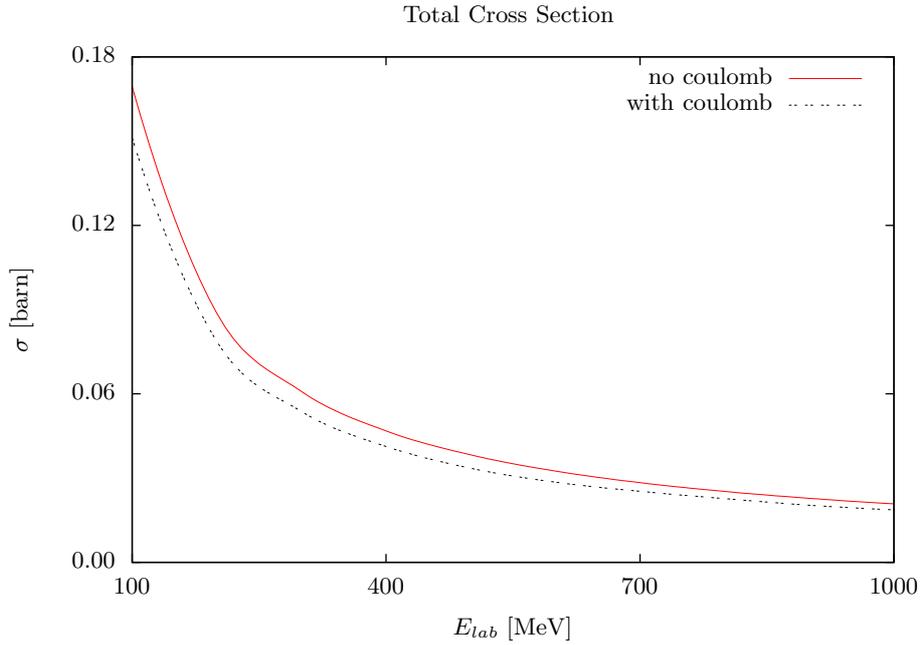}
\caption{(color online) Total cross section for various energies, with $\mu_C = 10$ MeV for
the Coulomb interaction.}
\label{fig2}
\end{figure}

Next we observ Coulomb effects in total cross section for various energies up to $E_{lab}=1000$ MeV. We take $\mu_C = 10$ MeV. 
These Coulomb effects are shown in Fig.~\ref{fig2}. Including Coulomb interaction lowers the total cross section, especially 
for lower energies. Varying energy to a high one together with using the 3D technique also leads to no difficulty, at least 
for the system being considered. 

\section{Summary}
\label{summary}
We have demonstrated the use of a 3D technique to calculate scattering of two bosons with Coulomb admixtures.
In observing some Coulomb effects we find it not difficult to vary the Coulomb range as well as the energy without 
much affecting the numerical parameters. For the system being considered $\mu_C=10$ MeV is sufficient to represent 
some kind of effective Coulomb range.


\end{document}